# ON THE VON NEUMANN AND SHANNON ENTROPIES FOR QUANTUM WALKS ON $Z^2$

CLEMENT AMPADU

*31 Carrolton Road*
*Boston, Massachusetts, 02132, USA*
*drampadu@hotmail.com*

We give asymptotic behaviors of the von Neumann entropy and the Shannon entropy of discrete-time time quantum walks on $Z^2$

### I. Introduction

The study of limit theorems for discrete-time quantum walks on various lattices has been the subject of extensive study by many authors, for examples see the review by Ampadu[1] and the references therein. The focus of this paper is on limit theorems for the von Neumann entropy and the Shannon entropy of the quantum walk on $Z^2$ starting from the origin with arbitrary coin and initial state.

Concerning the review on limiting values of these entropies, Carneiro et.al[2] studied the long-time asymptotic coin-position entanglement on various graphs by numerical simulations. Venegas-Andraca et.al[3] investigated the von Neumann entropy on $Z$ numerically. Using Fourier analysis techniques, asymptotic coin position entanglement of quantum walks for the Hadamard walk in the case of localized (setting of our paper) and non-localized conditions was analytically computed by Abal.et.al[4], on the other hand Annabestani et.al[5] gave an exact characterization of asymptotic coin position entanglement of quantum walks on $Z^2$. Liu et.al[6] presented limit theorems for the von Neumann entropy of quantum walks on the N-cycle. Bracken et.al[7] numerically computed the

Shannon entropy of the quantum walk on $Z$ defined by a coin which is a generalization of the Hadamard coin. Chandrashekar et.al[8] computed numerically the Shannon entropy for a generalization of the Hadamard walk in one dimension, denoted by $U(\theta)$ in this paper. For the symmetric random walk, Barron[9] and Takano[10] gave limiting values on the Shannon entropy in the likeness of Theorem 2 of this paper. Moreover they show that the rate of convergence of the entropy is related to the walkers position in the central limit theorem. In fact, we can infer from Theorem 2 of this paper that the rate of convergence of the Shannon entropy in the quantum case is also related to the central limit theorem, but the relation is slightly different from the random walk case.

Our motivation is as follows, recall that the von Neumann entropy can be used to measure entanglement between the coin and the particle position of quantum walks. Moreover, entanglement does not appear in classical systems, hence its importance for quantum information processing. On the other hand the Shannon entropy which is one of the basic quantities of information theory can be used to clarify the information included in the system. By studying the asymptotic behavior of these entropies we can control entanglement, and explain their rate of convergence via the distribution of the quantum walk.

The main results of this paper are contained in Theorems 1 and 2. In Theorem 1 we show limits of the von Neumann entropy and in Theorem 2 we show limits of the Shannon entropy. Using Theorem 1 we can explain the oscillatory behavior of entanglement and use the distribution of the quantum walk to explain the rate of convergence. We should remark that in the one dimensional setting Carneiro et.al[2] have numerically explained the oscillatory behavior of entanglement for quantum walks governed by $H = \begin{bmatrix} \sqrt{p} & \sqrt{1-p} \\ \sqrt{1-p} & -\sqrt{p} \end{bmatrix}$, which is a generalization of the standard Hadamard walk in one dimension. Theorem 2 on the other hand, as was mentioned earlier shows that the rate of convergence of the Shannon entropy is related to the central limit theorem. In the classical case, for the symmetric random walk, the limits of the Shannon entropy $S_n^{RW}$ [9,10] are given by $\lim_{n\to\infty} \dfrac{S_n^{RW}}{\log_2(\sqrt{n})} = 1$ and $\lim_{n\to\infty} \log_2(\sqrt{n})\left(\dfrac{S_n^{RW}}{\log_2(\sqrt{n})} - 1\right) = \dfrac{1}{2}\log_2(2\pi e)$

The rest of the paper is organized as follows. In Section II we give the definition of the quantum walk. Results on the von Neumann entropy are presented in Section III. In Section IV a long-time asymptotic for the Shannon entropy of the quantum walk is shown. Section V is devoted to the summary and an open problem.

**II. Definition**

Recall that the discrete-time quantum walk (QW) is the quantum analogue of the classical random walk with an additional degree of freedom called chirality. In the two dimensional setting the chirality takes values, left, right, downward, and upward, and means the direction of motion of the walker. At each time step, the particle moves according to its chirality state. For example, if the chirality state is upward, then the particle moves one step up. Let us define $|L\rangle = \begin{bmatrix} 1 \\ 0 \\ 0 \\ 0 \end{bmatrix}$, $|R\rangle = \begin{bmatrix} 0 \\ 1 \\ 0 \\ 0 \end{bmatrix}$, $|D\rangle = \begin{bmatrix} 0 \\ 0 \\ 1 \\ 0 \end{bmatrix}$, and $|U\rangle = \begin{bmatrix} 0 \\ 0 \\ 0 \\ 1 \end{bmatrix}$, where $L, R, D, U$ refer to the left, right, down, and up chirality states respectively. The time evolution of the quantum walk on $Z^2$ is determined by $U^{\otimes 2}$, where $U = \begin{bmatrix} a & b \\ c & d \end{bmatrix} \in U(2)$, with $a, b, c, d \in C$, where $C$ is the set of complex numbers. The unitarity of $U$ gives $|a|^2 + |b|^2 = |c|^2 + |d|^2 = 1$, $a\bar{c} + b\bar{d} = 0$, $c = -\Delta\bar{b}$, $d = \Delta\bar{a}$, where $\bar{z}$ denotes complex conjugation, and $\Delta = \det U = ad - bc$ with $|\Delta| = 1$. We should remark that $U^{\otimes 2}$ is a $4\times 4$ matrix which is also unitary. Let us write $U(\theta)^{\otimes 2}$, where $U(\theta) = \begin{bmatrix} \cos\theta & \sin\theta \\ \sin\theta & -\cos\theta \end{bmatrix}$ and $0 < \theta < \frac{\pi}{2}$.

We should remark that a generalization of $U(\theta)^{\otimes 2}$ was studied by Ampadu[11], namely $H^{**}(p,q)$ in his paper. Moreover, $U\left(\frac{\pi}{4}\right)^{\otimes 2} = H^{**}\left(\frac{1}{2}, \frac{1}{2}\right)$, that is we get the standard two dimensional Hadamard walk. In order to define the dynamics of the model, we write

$$U^{\otimes 2} = P + Q + R + S$$

where $P = \begin{bmatrix} a^2 & ab & ab & b^2 \\ 0 & 0 & 0 & 0 \\ 0 & 0 & 0 & 0 \\ 0 & 0 & 0 & 0 \end{bmatrix}$, $Q = \begin{bmatrix} 0 & 0 & 0 & 0 \\ ac & ad & bc & bd \\ 0 & 0 & 0 & 0 \\ 0 & 0 & 0 & 0 \end{bmatrix}$, $R = \begin{bmatrix} 0 & 0 & 0 & 0 \\ 0 & 0 & 0 & 0 \\ ac & bc & ad & bd \\ 0 & 0 & 0 & 0 \end{bmatrix}$, and

$S = \begin{bmatrix} 0 & 0 & 0 & 0 \\ 0 & 0 & 0 & 0 \\ 0 & 0 & 0 & 0 \\ c^2 & cd & cd & d^2 \end{bmatrix}$. The matrices $P, Q, R, S$ represent the walker moving to the left, right, down, and up directions, respectively, at each time step. Let $\Xi_n(l, r, d, u)$ denote the sum of all paths starting from the origin in the trajectory consisting of $l$ steps left, $r$ steps right, $d$ steps downwards, and $u$ steps upwards. For time $n = l + r + d + u$, and position $x = -l + r$, $y = -d + u$, we have

$$\Xi_n(l,r,d,u) = \sum_{l_j, r_j, d_j, u_j} P^{l_1} Q^{r_1} R^{d_1} S^{u_1} \cdots P^{l_{n-1}} Q^{r_{n-1}} R^{d_{n-1}} S^{u_{n-1}} P^{l_n} Q^{r_n} R^{d_n} S^{u_n}$$, where the summation is taken

over all integers $l_j, r_j, d_j, u_j \geq 0$ satisfying $\sum_{i=1}^n l_i = l$, $\sum_{i=1}^n r_i = r$, $\sum_{i=1}^n d_i = d$, $\sum_{i=1}^n u_i = u$, and $l_j + r_j + d_j + u_j = 1$. We should note that the definition implies we can write

$$\Xi_{n+1}(l,r,d,u) = P\Xi_n(l-1,r,d,u) + Q\Xi_n(l,r-1,d,u) + R\Xi_n(l,r,d-1,u) + S\Xi_n(l,r,d,u-1).$$ Let

the set of initial qubit states at the origin for the quantum walk be given by

$$\Phi = \{\varphi = \alpha|L\rangle + \beta|R\rangle + \gamma|D\rangle + \lambda|U\rangle \in C^4 : |\alpha|^2 + |\beta|^2 + |\gamma|^2 + |\lambda|^2 = 1\}$$

$$= \{\varphi =^T [\alpha \quad \beta \quad \gamma \quad \lambda] \in C^4 : |\alpha|^2 + |\beta|^2 + |\gamma|^2 + |\lambda|^2 = 1\}$$

The probability that the quantum walker is in position $(x, y)$ at time $n$ starting from the origin with $\varphi \in \Phi$ is defined by $P(X_n = x, Y_n = y) = \|\Xi(l,r,d,u)\varphi\|^2$, where $n = l+r+d+u$, $x = -l+r$, and $y = -d+u$. The probability amplitude $\Psi_n(x, y)$ at position $(x, y)$ at time $n$ is given by

$$\Psi_n(x,y) = \sum_{j \in \{L,R,D,U\}} \Psi_n^j(x,y) |j\rangle = \Xi_n(l,r,d,u)\varphi. \text{ So, } P(X_n = x, Y_n = y) = \sum_{j \in \{L,R,D,U\}} |\Psi_n^j(x,y)|^2. \text{ Let}$$

$|\Psi_n\rangle =^T [\cdots, \Psi(-1,0), \Psi(0,0), \Psi(1,0), \cdots]$. The density operator at time $n$ is defined by $\rho_n = |\Psi_n\rangle\langle\Psi_n|$.

Entanglement for pure states can be quantified by the von Neumann entropy of the reduced density operator which is defined by $\rho_n^c = Tr_p(\rho_n)$, where the partial trace is taken over position. The associated von Neumann entropy at time $n$ is defined by $S_n = -Tr\{\rho_n^c \log_2(\rho_n^c)\}$, which is known as the entropy of entanglement, and quantifies the quantum correlations present in the pure state. For

$i, j \in \{L, R, D, U\}$, let $\|\Psi_n^j\|^2 = \sum_{x,y=-n}^{n} |\Psi_n^j(x,y)|^2$ and for $i \neq j$ let $\langle \Psi_n^i, \Psi_n^j \rangle = \sum_{x,y=-n}^{n} \Psi_n^i(x,y)^* \Psi_n^j(x,y)$.

Note that $\sum_{j \in \{L,R,D,U\}} \|\Psi_n^j\|^2 = 1$ for any time $n$. The entropy of entanglement can be obtained by diagonalization of $\rho_n^c$. This is represented by the following $4 \times 4$ Hermitian matrix,

$$\rho_n^c = \begin{bmatrix} \|\Psi_n^L\|^2 & \langle\Psi_n^R,\Psi_n^L\rangle & \langle\Psi_n^D,\Psi_n^L\rangle & \langle\Psi_n^U,\Psi_n^L\rangle \\ \langle\Psi_n^L,\Psi_n^R\rangle & \|\Psi_n^R\|^2 & \langle\Psi_n^D,\Psi_n^R\rangle & \langle\Psi_n^U,\Psi_n^R\rangle \\ \langle\Psi_n^L,\Psi_n^D\rangle & \langle\Psi_n^R,\Psi_n^D\rangle & \|\Psi_n^D\|^2 & \langle\Psi_n^U,\Psi_n^D\rangle \\ \langle\Psi_n^L,\Psi_n^U\rangle & \langle\Psi_n^R,\Psi_n^U\rangle & \langle\Psi_n^D,\Psi_n^U\rangle & \|\Psi_n^U\|^2 \end{bmatrix}. \text{ Let us write } |L_1\rangle = \begin{bmatrix} 1 \\ 0 \end{bmatrix} \text{ and } |L_2\rangle = \begin{bmatrix} 0 \\ 1 \end{bmatrix} \text{ to}$$

correspond to the chirality states left and right in the one dimensional setting, then it is easily seen that

$|L_1\rangle \otimes |L_2\rangle = |R\rangle$, $|L_1\rangle \otimes |L_1\rangle = |L\rangle$, $|L_2\rangle \otimes |L_1\rangle = |D\rangle$, and $|L_2\rangle \otimes |L_2\rangle = |U\rangle$. So the matrix

$$\begin{bmatrix} \|\tilde{\Psi}_n^L\|^2 & \langle \tilde{\Psi}_n^R, \tilde{\Psi}_n^L \rangle \\ \langle \tilde{\Psi}_n^L, \tilde{\Psi}_n^R \rangle & \|\tilde{\Psi}_n^R\|^2 \end{bmatrix} \otimes \begin{bmatrix} \|\tilde{\Psi}_n^L\|^2 & \langle \tilde{\Psi}_n^R, \tilde{\Psi}_n^L \rangle \\ \langle \tilde{\Psi}_n^L, \tilde{\Psi}_n^R \rangle & \|\tilde{\Psi}_n^R\|^2 \end{bmatrix}$$ is in a one-to-one correspondence with the

matrix $\rho_n^c$ in this paper. In Ide et.al[12] it is shown that the matrix $\begin{bmatrix} \|\tilde{\Psi}_n^L\|^2 & \langle \tilde{\Psi}_n^R, \tilde{\Psi}_n^L \rangle \\ \langle \tilde{\Psi}_n^L, \tilde{\Psi}_n^R \rangle & \|\tilde{\Psi}_n^R\|^2 \end{bmatrix}$ has eigenvalues $r_{n,\pm} = \dfrac{1 \pm \sqrt{1 - 4\Delta_n(\rho)}}{2}$, where $\Delta_n(\rho)$ is

the determinant of $\begin{bmatrix} \|\tilde{\Psi}_n^L\|^2 & \langle \tilde{\Psi}_n^R, \tilde{\Psi}_n^L \rangle \\ \langle \tilde{\Psi}_n^L, \tilde{\Psi}_n^R \rangle & \|\tilde{\Psi}_n^R\|^2 \end{bmatrix}$. Since the tensor product decomposition is shown to be

in a one-to-onecorrespondence with the matrix $\rho_n^c$, it follows that the eigenvalues of $\rho_n^c$ is the product of the eigenvalues of $\begin{bmatrix} \|\tilde{\Psi}_n^L\|^2 & \langle \tilde{\Psi}_n^R, \tilde{\Psi}_n^L \rangle \\ \langle \tilde{\Psi}_n^L, \tilde{\Psi}_n^R \rangle & \|\tilde{\Psi}_n^R\|^2 \end{bmatrix}$. In particular from $r_{n,\pm} = \dfrac{1 \pm \sqrt{1 - 4\Delta_n(\rho)}}{2}$, the

eigenvalues of $\rho_n^c$ are given by $\lambda_1(\rho_n^c) = r_{n,+}^2$, $\lambda_2(\rho_n^c) = r_{n,-}^2$, $\lambda_3(\rho_n^c) = \lambda_4(\rho_n^c) = r_{n,-} r_{n,+}$. So the

reduced entropy can be expressed as $S_n^c = -\left( \sum_{j=1}^{4} \lambda_j(\rho_n^c) \log_2(\lambda_j(\rho_n^c)) \right)$, where $\lambda_j(\rho_n^c)$ are the

eigenvalues of $\rho_n^c$. On the other hand, the entropy of the reduced density matrix of the position at

time $n$ is defined by $\rho_n^p = Tr_c(\rho_n)$, which quantifies the entanglement between the coin and the

walker's position. This is represented by a Hermitian matrix whose $((x,y),z)$ element is given by

$\rho_n^p((x,y),z) = \sum_{j \in \{L,R,D,U\}} \Psi_n^j(x,y) \Psi_n^j(z)^*$, where $x,y,z \in Z$. If $z = (x,y)$, then the diagonal elements

become $\sum_{j \in \{L,R,D,U\}} \Psi_n^j(x,y) \Psi_n^j(x,y)^* = P(X_n = x, Y_n = y)$. If $|z| > n$, then $\rho_n^p((x,y),z) = 0$, then the reduced entropy $S_n^p$ can be written as $S_n^p = -\sum_{x,y=-n}^{n} r_n(x,y) \log(r_n(x,y))$ where $r_n(x,y)$ are the eigenvalues of a Hermitian matrix with the element $\rho_n^p((x,y),z)$ for $x,y,z \in \{-n, -n+1, \cdots, n\}$. For

$n \geq 0$, $S_n^c = S_n^P$, so we focus on $S_n^c$.

### III. Main Results on Entropy of Entanglement

In this section we present the results for asymptotic coin-position entanglement for the quantum walk on $Z^2$. We should remark that the proof of Theorem 1 follows the argument in Konno[13-14], and in particular the proof of Theorem 3.1 in Ide.et al[12], therefore we omit it. For $i, j \in \{L, R, D, U\}$, put

$$\|\Psi_\infty^j\| = \lim_{n \to \infty} \|\Psi_n^j\| \text{ and for } i \neq j \text{ put } \langle \Psi_\infty^i, \Psi_\infty^j \rangle = \lim_{n \to \infty} \langle \Psi_n^i, \Psi_n^j \rangle, \text{ then we have the following.}$$

**Theorem 1:** *When the quantum walk is determined by $U^{\otimes 2}$ with $abcd \neq 0$, we have for*

$$i, j \in \{L, R, D, U\}, \quad \|\Psi_\infty^i\|^2 = \frac{1}{4\pi^2} \int_{-|d|^2}^{|d|^2} \int_{-|a|^2}^{|a|^2} \frac{h^i(x, y)}{\sqrt{|a|^4 - x^2} \sqrt{|d|^2 - y^2}} dxdy, \text{ and for } i \neq j$$

$$\langle \Psi_\infty^i, \Psi_\infty^j \rangle = \frac{1}{4\pi^2} \int_{-|d|^2}^{|d|^2} \int_{-|a|^2}^{|a|^2} \frac{h^{ij}(x, y)}{\sqrt{|a|^4 - x^2} \sqrt{|d|^2 - y^2}} dxdy, \text{ where } h^i(x, y) \text{ for } i \in \{L, R, D, U\} \text{ is given by}$$

$$h^L(x, y) = \frac{A}{|a|^2} \left(\frac{1-x}{1+x}\right)^{\frac{1}{2}} \left(\frac{1-y}{1+y}\right)^{\frac{1}{2}} + \frac{|D|}{|C^2|} \left(\frac{1}{1+x}\right)^{\frac{1}{2}} \left(\frac{1}{1+y}\right)^{\frac{1}{2}} + 2 \operatorname{Re}\left(\frac{1-\gamma}{a^2\gamma} \frac{AD^*}{(c^2)^*}\right)$$

$$h^R(x, y) = \frac{|B|^2}{|ad|^2} \left(\frac{1-x}{1+x}\right)^{\frac{1}{2}} \left(\frac{1-y}{1+y}\right)^{\frac{1}{2}} + \frac{|C|}{|bc|^2} \left(\frac{1}{1+x}\right)^{\frac{1}{2}} \left(\frac{1}{1+y}\right)^{\frac{1}{2}} + 2 \operatorname{Re}\left(\frac{r-\gamma}{ad\gamma} \frac{BC^*}{bc^*}\right)$$

$$h^D(x, y) = \frac{|C|^2}{|ad|^2} \left(\frac{1-x}{1+x}\right)^{\frac{1}{2}} \left(\frac{1-y}{1+y}\right)^{\frac{1}{2}} + \frac{|B|^2}{|bc|^2} \left(\frac{1}{1+x}\right)^{\frac{1}{2}} \left(\frac{1}{1+y}\right)^{\frac{1}{2}} + 2 \operatorname{Re}\left(\frac{d-\gamma}{ad\gamma} \frac{(BC)^*}{(bc)^*}\right)$$

$$h^U(x, y) = \frac{|D|^2}{|d|^2} \left(\frac{1-x}{1+x}\right)^{\frac{1}{2}} \left(\frac{1-y}{1+y}\right)^{\frac{1}{2}} + \frac{|A|^2}{|b^2|^2} \left(\frac{1}{1+x}\right)^{\frac{1}{2}} \left(\frac{1}{1+y}\right)^{\frac{1}{2}} + 2 \operatorname{Re}\left(\frac{u-\gamma}{d^2\gamma} \frac{DA^*}{(b^*)^2}\right), \text{ where}$$

$A = a^2\alpha + ab(\beta + \gamma) + b^2\lambda$, $B = (ac)\alpha + (bc)\beta + (ad)\gamma + (bd)\lambda$,

$C = (ac)\alpha + (bc)\beta + (ad)\gamma + (bd)\lambda$, and $D = c^2\alpha + cd(\beta + \gamma) + d^2\lambda$, and $h^{ij}(x, y)$ is determined

*in a similar way.* Then

$$\lim_{n \to \infty} S_n^c = -\{r_{\infty,+}^2 \log_2(r_{\infty,+}^2) + r_{\infty,-}^2 \log_2(r_{\infty,-}^2) + 2r_{\infty,+}r_{\infty,-} \log_2(r_{\infty,+}r_{\infty,-})\}, \text{ where } r_{\infty,\pm}^2 = \lim_{n \to \infty} r_{n,\pm}^2,$$

$r_{\infty,+}r_{\infty,-} = \lim_{n \to \infty} r_{n,+}r_{n,-}$, and $r_{n,\pm}^2$ and $r_{n,+}r_{n,-}$ *(multiplicity 2) are the eigenvalues of* $\rho_n^c$.

We should remark for $U(\theta)^{\otimes 2}$, where $U(\theta) = \begin{bmatrix} \cos\theta & \sin\theta \\ \sin\theta & -\cos\theta \end{bmatrix}$ and $0 < \theta < \frac{\pi}{2}$, we get similar integral representations for a two dimensional generalization of the Hadamard walk, in particular the Corollary to Theorem 1 is obtained by taking $a = -d = \cos\theta$ in the double integrals.

### IV. Main Results on Shannon Entropy

In this section we obtain the limit theorems associated with the Shannon entropy of the quantum walk. Let $P(x_n, y_n) = P(X_n = x, Y_n = y)$. Define the Shannon entropy of the quantum walk by

$$S_n = -\sum_{x,y=-n}^{n} P_n(x,y) \log_2(P_n(x,y)). \text{ For } W \in \{L, R, D, U\}, \text{ put } h(x,y) = \frac{|\Psi_n^W(x,y)|^2}{\|\Psi_n^W\|^2}, \text{ and let}$$

$$S_n^W = -\sum_{x,y=-n}^{n} h(x,y) \log_2(h(x,y)), \text{ define } \rho^W = \lim_{n \to \infty} \|\Psi_n^W\|^2, \ f^W(x,y) = \frac{h^W(x,y)}{\sqrt{|a|^4 - x^2}\sqrt{|d|^4 - y^2}}, \text{ and}$$

moreover, $f(x,y) = \sum_{W \in \{L,R,D,U\}} f^W(x,y)$. We give the limit theorems for $S_n^j$, $j \in \{L, R, D, U\}$, and

$S_n$. In what follows we will derive the result in the case of $S_n^L$, as the others follow in a similar way.

Notice from the definition of $S_n^L$ we have the following

$$S_n^L = \log_2\left(\|\Psi_n^L\|^2\right) - \frac{1}{\|\Psi_n^L\|^2} \sum_{x,y=-n}^{n} |\Psi_n^L(x,y)|^2 \log_2\left(|\Psi_n^L(x,y)|^2\right)$$

Since $\rho^L = \lim_{n \to \infty} \|\Psi_n^L\|^2$, we have $\lim_{n \to \infty} S_n^L = \log_2(\rho^L) - \frac{1}{\rho^L} \sum_{x,y=-n}^{n} |\Psi_n^L(x,y)|^2 \log_2\left(|\Psi_n^L(x,y)|^2\right)$.

We should remark that Theorem 1 can also be obtained using asymptotics concerning the Jacobi polynomials and the Riemann Lebesgue lemma. In a similar way we have the following asymptotics

$$\sum_{x,y=-n}^{n} |\Psi_n^L(x,y)|^2 \log_2\left(|\Psi_n^L(x,y)|^2\right)$$

$$\sim \int_{-|d|^2}^{|d|^2}\int_{-|a|^2}^{|a|^2} f^L(x,y)\log_2\left(\frac{4f^L(x,y)}{n}\right)dxdy$$

$$= -\rho^L \times \log_2\left(\frac{n}{4}\right) + \int_{-|d|^2}^{|d|^2}\int_{-|a|^2}^{|a|^2} f^L(x,y)\left(\log_2\left(f^L(x,y)\right)\right)dxdy$$

So, $S_n^L \sim \log_2\left(\frac{n}{4}\right) - \int_{-|d|^2}^{|d|^2}\int_{-|a|^2}^{|a|^2} \frac{f^L(x,y)}{\rho^L}\left(\log_2\left(\frac{f^L(x,y)}{\rho^L}\right)\right)dxdy$. In particular, we have the following theorem.

**Theorem 2**: If the quantum walk is determined by $U^{\otimes 2}$ with $abcd \neq 0$, then

$$\lim_{n\to\infty}\frac{S_n^W}{\log_2\left(\frac{n}{4}\right)} = \lim_{n\to\infty}\frac{S_n}{\log_2\left(\frac{n}{4}\right)} = 1, \text{ where } W \in \{L,R,D,U\}. \text{ Further,}$$

$$\lim_{n\to\infty}\log_2\left(\frac{n}{4}\right)\left(\frac{S_n^W}{\log_2\left(\frac{n}{4}\right)} - 1\right) = \int_{-|d|^2}^{|d|^2}\int_{-|a|^2}^{|a|^2} \frac{f^W(x,y)}{\rho^W}\left(\log_2\left(\frac{f^W(x,y)}{\rho^W}\right)\right)dxdy$$

$$\lim_{n\to\infty}\log_2\left(\frac{n}{4}\right)\left(\frac{S_n}{\log_2\left(\frac{n}{4}\right)} - 1\right) = \int_{-|d|^2}^{|d|^2}\int_{-|a|^2}^{|a|^2} f(x,y)\left(\log_2\left(f(x,y)\right)\right)dxdy$$

where $W \in \{L,R,D,U\}$.

## V. Summary and Open Problem

In this paper, we have obtained limit theorems for the von Neumann entropy and the Shannon entropy for quantum walks on $Z^2$ starting from the origin with arbitrary coin and initial state. It is an open problem to generalize the path counting technique for quantum walks on $d-$dimensional lattices, say, $Z^d$, and use it to give general analytic formula for the limiting values of the entropies explicitly. This problem is extreme, in the sense that the combinatorics surrounding the path counting technique on higher dimensional lattices is not well understood.